\documentclass[aps,showpacs,twocolumn,preprint,amsmath,amssymb,superscriptaddress,intlimits]{revtex4-1}

\usepackage{graphicx}
\usepackage{subfigure}
\usepackage{color}
\usepackage{epstopdf}
\usepackage{amssymb}
\usepackage{amsmath}
\usepackage{amsfonts}
\usepackage{dcolumn}
\usepackage{bm}
\usepackage{soul}
\usepackage{array}
\usepackage{physics}

\newcommand{\vk}{{\bf k}}
\newcommand{\vQ}{{\bf Q}}
\newcommand{\vq}{{\bf q}}

\newcolumntype{C}[1]{>{\centering\let\newline\\\arraybackslash\hspace{0pt}}m{#1}}

\graphicspath{{figures/}}

\def\correspondingauthor{\footnote{sglouie@berkeley.edu}}

\begin{document}

\title{Exciton lifetime and optical linewidth profile via exciton-phonon interactions: Theory and first-principles calculations for monolayer MoS$_2$}
\author{Y.-H. Chan}
\affiliation{Institute of Atomic and Molecular Sciences, Academia Sinica, and Physics Division, National Center of Theoretical Physics, Taipei 10617, Taiwan}
\affiliation{Department of Physics, University of California, Berkeley, CA, 94720-7300, USA}
\affiliation{Materials Science Division, Lawrence Berkeley National Laboratory, Berkeley, CA, 94720, USA}

\author{Jonah B. Haber}
\affiliation{Department of Physics, University of California, Berkeley, CA, 94720-7300, USA}
\affiliation{Materials Science Division, Lawrence Berkeley National Laboratory, Berkeley, CA, 94720, USA}

\author{Mit H. Naik}
\affiliation{Department of Physics, University of California, Berkeley, CA, 94720-7300, USA}
\affiliation{Materials Science Division, Lawrence Berkeley National Laboratory, Berkeley, CA, 94720, USA}

\author{J. B. Neaton}
\affiliation{Department of Physics, University of California, Berkeley, CA, 94720-7300, USA}
\affiliation{Materials Science Division, Lawrence Berkeley National Laboratory, Berkeley, CA, 94720, USA}
\affiliation{Kavli Energy NanoSciences Institute at Berkeley, University of California, Berkeley, California 94720, USA}

\author{Diana Y. Qiu}
\affiliation{Department of Mechanical Engineering and Materials Science, Yale University, New Haven, CT 06520}

\author{Felipe H. da Jornada}
\affiliation{Department of Materials Science and Engineering, Stanford University, Stanford, CA 94305, USA}
\affiliation{Stanford PULSE Institute, SLAC National Accelerator Laboratory, Menlo Park, California 94025, USA}

\author{Steven G. Louie \correspondingauthor{}}
\affiliation{Department of Physics, University of California, Berkeley, CA, 94720-7300, USA}
\affiliation{Materials Science Division, Lawrence Berkeley National Laboratory, Berkeley, CA, 94720, USA}

\begin{abstract}

Exciton dynamics dictate the evolution of photoexcited carriers in photovoltaic and optoelectronic devices. However, interpreting their experimental signatures is a challenging theoretical problem due to the presence of both electron-phonon and many-electron interactions. We develop and apply here a first-principles approach to exciton dynamics resulting from exciton-phonon coupling in monolayer MoS$_2$ and reveal the highly selective nature of exciton-phonon coupling due to the internal spin structure of excitons, which leads to a surprisingly long lifetime of the lowest energy bright A exciton. Moreover, we show that optical absorption processes rigorously require a second-order perturbation theory approach, with photon and phonon treated on an equal footing, as proposed by Toyozawa and Hopfield.
Such a treatment, thus far neglected in first-principles studies, gives rise to off-diagonal exciton-phonon coupling matrix elements, which are critical for the description of dephasing mechanisms, and yields exciton linewidths in excellent agreement with experiment.

\end{abstract}

\date{\today}
\maketitle

In low-dimensional or nanostructured semiconductors, the low-energy optical excitations are dominated by strongly bound correlated electron-hole pairs known as excitons. Understanding exciton energetics and dynamics is essential for diverse applications across optoelectronics, quantum information and sensing, as well as energy harvesting and conversion. By now, it is well-established that these large excitonic effects are a combined consequence of quantum confinement and reduced screening in low dimensions~\cite{Qiu2013, Qiu2016, Chernikov2014, Wang2018}. However, many challenges remain in understanding the dynamics of these excitons, especially when it comes to correlating complex experimental signatures observed in various spectroscopies with underlying dynamical processes through the use of quantitatively predictive \textit{ab initio} theories.

Exciton-phonon interactions play a key role in determining exciton nonradiative dynamics, decoherence times, temperature-dependent behavior, linewidths, and diffusion.
Experimentally, exciton-phonon interaction are embedded in various spectroscopic measurements including absorption, photoluminescence, and four-wave mixing measurements. For example, the exciton-phonon coupling strength has been inferred for various transition metal dichalogenides (TMDs), such as WSe$_2$ and MoS$_2$, based on the spectroscopic linewidth or the evolution of exciton spectral intensity in pump-probe setups. However, the reported lifetimes vary by an order of magnitude \cite{Moody2015,Dey2016,Cadiz2017} as a consequence of variations in substrate and sample preparations, highlighting the need for accurate first-principles theories.

On the theoretical side, \textit{ab initio} methods for determining electron-phonon interactions using density-functional perturbation theory (DFPT) are by now well-established~\cite{Baroni2001,Gianozzi1991}. Inclusion of self-energy corrections to the electron-phonon interaction for more correlated systems may also be included at the GW level using a recently developed GW perturbation theory (GWPT)~\cite{Li2019}. In the limit where the exciton binding energy is small, one can approximate the exciton-phonon renormalization of the exciton energy and linewidth in terms of the electron-phonon renormalization for the constituent electron states and hole states of each exciton state~\cite{Marini2008}, and it has been demonstrated that such a simplified approach is nonetheless essential for reproducing the experimental line-shape in optical spectra of monolayer MoS$_2$~\cite{Qiu2013,Molina2016}. At a higher level of theory, one can also derive the exciton-phonon coupling matrix elements and the exciton self-energy within many-body perturbation theory~\cite{Toyozawa1958,Hopfield1961,Toyozawa1964,Segall1968,Jiang2007,Antonius2017,Shree2018,Chen2020}, and the diagonal part of the exciton self-energy due to exciton-phonon coupling has been recently applied to study exciton-phonon interactions in bulk hexagonal boron nitride~\cite{Cannuccia2019,Chen2020}. However, model calculations based on the semiconductor Bloch equations suggest that the off-diagonals of the exciton self energy due to exciton-phonon coupling in the original unperturbed exciton basis (see definition below), which have thus far been overlooked in \textit{ab initio} calculations, can play an important role in the exciton linewidth~\cite{Selig2016,Christiansen2017,Brem2018,Brem2019}. The full phase space in \textit{ab initio} calculations of exciton-phonon interactions in low-dimensional materials, where exciton effects are strongest, has also not yet been fully explored, with existing studies limited to either frozen phonons~\cite{huang2021} or $\Gamma$-point only calculations~\cite{Reichardt2020} and a few recent works based on many-body perturbation theory~\cite{Zhang2021,Zhang2022}.


\begin{figure}[t]
     \centering
     \includegraphics[width=.48\textwidth]{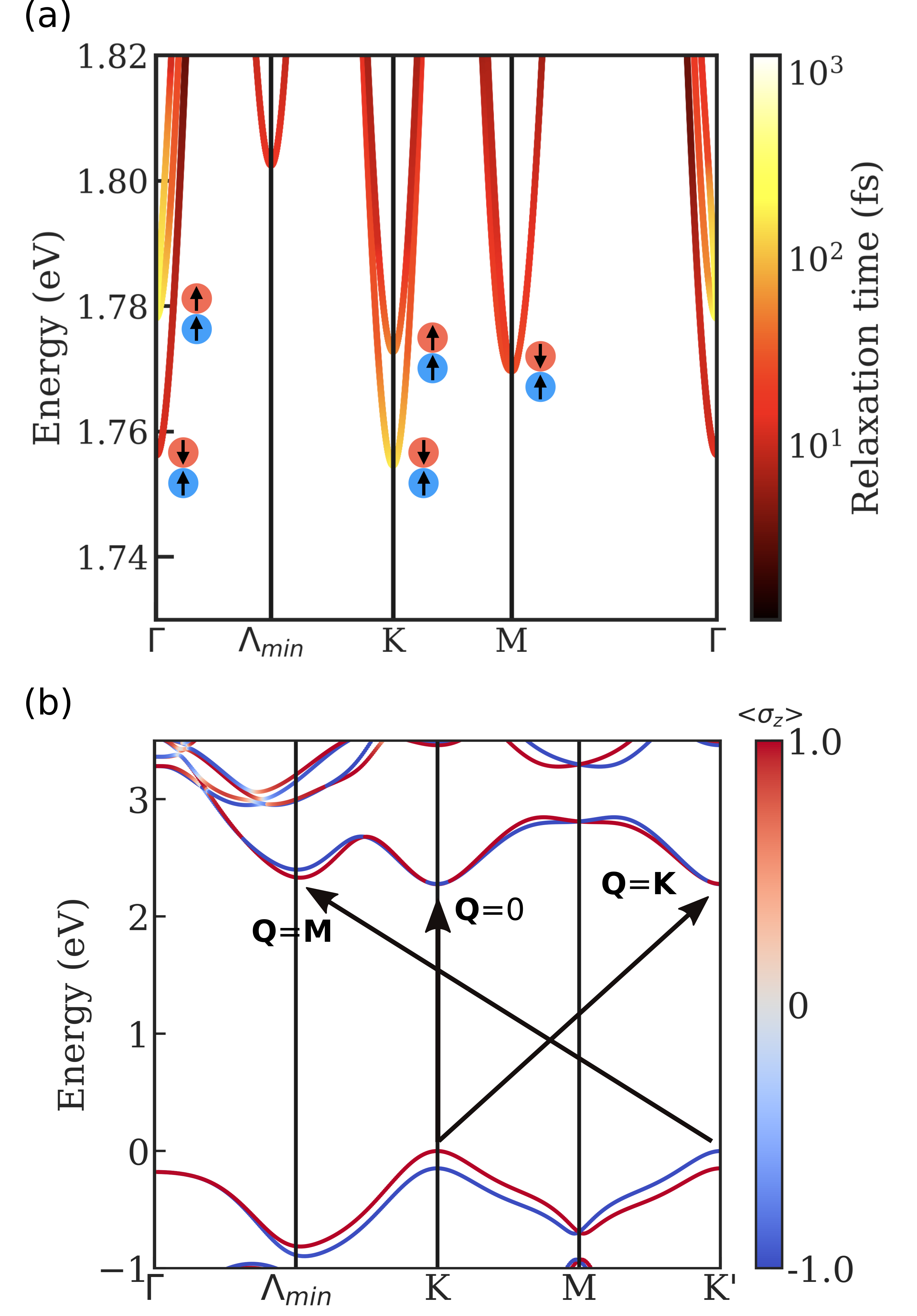}
\caption{(a) Exciton band structure of monolayer MoS$_2$ including both opposite-spin and parallel-spin states. Color indicates the relaxation time due to exciton-phonon coupling at 300K. Dots with arrows indicate the internal spin structure of excitons at valleys. (b) GW electron band structure with a color map of spin expectation values. Energy transition corresponding to excitons with center of mass momentum $\mathbf{Q=0}$, $\mathbf{Q=K}$, and $\mathbf{Q=M}$ are shown with the labeled arrows. The other $\mathbf{Q=M}$ excitons with the transition from $\mathbf{K'}$ valley valence electron to $\Lambda_{min}$ conduction electron is not shown.}
\label{fig:Relax}
\end{figure}

Motivated by the great interests in exciton dynamics and the lack of a predictive \textit{ab initio} theory, we develop an \textit{ab initio} theory to study exciton-phonon coupling in optical spectroscopy and apply it to compute both exciton relaxation and the lineshape of the absorption spectrum in monolayer MoS$_2$, a prototypical quasi two-dimensional (quasi-2D) material. Using a fully relativistic formalism, we reveal that the population lifetime or relaxation time of the lowest-energy bright A exciton due to exciton-phonon interactions is surprisingly longer than that of the lower-energy parallel-spin states at room temperature. This counterintuitive result arises as a consequence of the selection rules  of the exciton-phonon coupling, which do not allow for direct intervalley scattering from the A state. We show that phonon-mediated intervalley exciton scattering must be facilitated through other valleys in the conduction band~\cite{Li2013,Guo2019}. Additionally, we find that the absorption spectrum of the A exciton features an asymmetric lineshape due to interference effects from transitions between distinct exciton bands, which cannot be understood in the context of previous \textit{ab initio} theories. The computed linewidth and energy shift in the absorption spectrum with interference effects are in excellent agreement with experimental results. 

We first analyze the exciton dispersion of MoS$_2$ via the \textit{ab initio} GW plus Bethe-Salpeter equation (GW-BSE) method~\cite{Hybertsen1986,Rohlfing2000} by solving for excitons with finite center-of-mass (COM) momentum $\vQ$ with the BerkeleyGW package~\cite{Deslippe2012,Qiu2015}. Fig.~\ref{fig:Relax} (a) shows the exciton dispersion along a high symmetry path in the Brillouin zone. The lowest energy exciton states at $\Gamma$ ($\mathbf{Q}=0$) are doubly degenerate parallel-spin ($S^{\rm tot}=1$) optically dark excitons. We use the total spin $S^{\rm tot}$ to label exciton states, which is approximately a good quantum number in the K valleys even in the presence of spin-orbit coupling (SOC). Two bright exciton states, corresponding to the measured `A' peak in the absorption spectrum, are about 20 meV higher in energy. These bright exciton bands are doubly degenerate at $\Gamma$, with one band dispersing linearly and one band dispersing parabolically, consistent with previous calculations~\cite{Qiu2015,Qiu2021}. The lowest-energy exciton with $\vQ=K$ consists of electron and hole states from the $K$ and $K'$ valleys, vice versa, (Fig.~\ref{fig:Relax} (b)) and is a parallel-spin state. In contrast, the second-lowest at $\vQ=K$ exciton is a opposite-spin ($S^{\rm tot}=0$) state. The lowest energy $\vQ=M$ exciton also is a parallel-spin state -- composed for example of a hole in the $K'$ valley and an electron in the conduction band valley along the $\Lambda$ high-symmetry line (i.e., the $\Lambda_\mathrm{min}$ valley), while the second lowest-energy exciton is a opposite-spin state. This well-defined internal spin-texture of the excitons, leads to strict selection rules for exciton-phonon coupling that significantly extend the lifetime of the opposite-spin states at $\mathbf{Q}=0$, as we discuss later. 

Exciton-phonon coupling matrix elements $G_{SS'\nu}$ can be written as a superposition of electron-phonon coupling matrix elements $g_{nm\nu}$ weighted by the exciton envelope function (when expanded in terms of interband electron-hole pairs in $\mathbf{k}$-space) as derived in previous works~\cite{Toyozawa1958,Antonius2017,Chen2020}. The lowest-order exciton self-energy expression due to exciton-phonon coupling derived from many-body perturbation theory is similar to the Fan-Migdal term in the electron self-energy due to electron-phonon coupling and reads 
\begin{align}
&\Sigma_{S\vQ}(\omega) \nonumber \\
&= \frac{1}{N_\mathbf{q}}\sum_{S',\vq,\nu,\pm}\frac{|G_{S'S\nu}(\vQ,\vq)|^2(N_{\nu\vq} +\frac{1}{2}\pm\frac{1}{2})}{\omega-E_{S'\vQ+\vq}\mp\omega_{\nu\vq}+i\eta},
\label{Eq:sediag}
\end{align}
where $E_{S'\vQ+\vq}$ is the exciton energy of a state with exciton band index $S'$  and a COM momentum $\vQ+\vq$; $\omega_{\nu\vq}$ and $N_{\nu \vq}$ are the phonon frequency and Bose-Einstein occupation factor with $\nu$ labeling the phonon branch index and $\mathbf{q}$ the crystal momentum;  $N_{\mathbf{q}}$ is the number of the wavevectors sampled in the Brillouin zone; and finally, $G_{S'S\nu}(\vQ,\vq)$ is the exciton-phonon coupling matrix element encoding the probability amplitude for an exciton initially in state ($S, \vQ$) to scatter to state ($S', \vQ+\vq$) through the emission or absorption of a phonon $(\nu, \pm \vq)$.  The real part of Eq.~\ref{Eq:sediag} determines exciton energy renormalization while the imaginary part is proportional to the relaxation time of excitons. Eq.~\ref{Eq:sediag}, as we shall present below, needs to be extended to a matrix form to fully describe the coupled system.

\begin{figure*}[t]
     \centering
     \includegraphics[width=\textwidth]{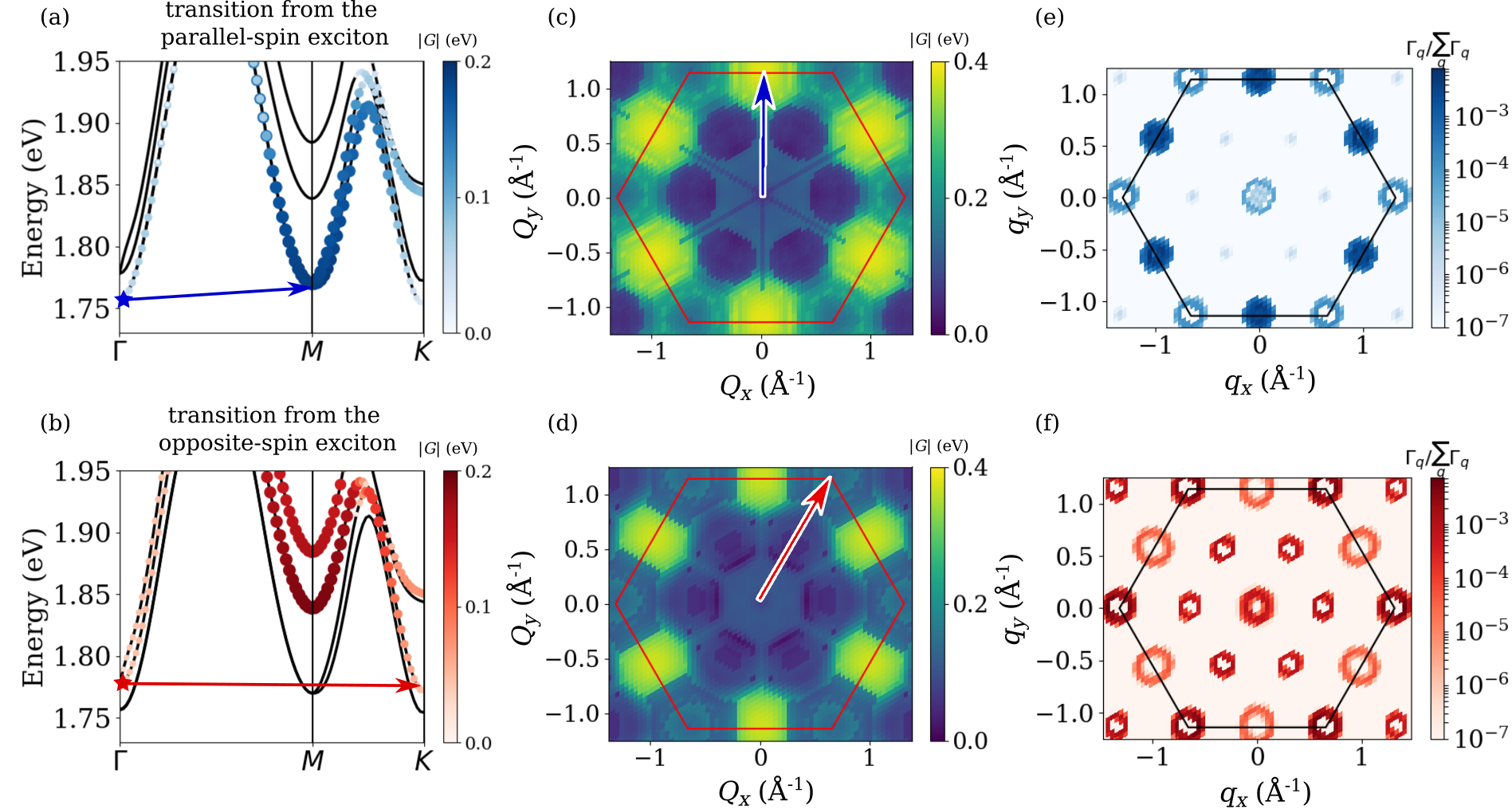}
\caption{Exciton-phonon coupling strength and phonon momentum \vq-resolved contribution to the total scattering rate for the lowest parallel-spin (blue star and panel (a),(c),(e)) and opposite-spin (red star and panel (b),(d),(f)) exciton state with a center of mass momentum $Q=\Gamma$. The color map and symbol sizes in (a) and (b) characterize the overall coupling strength from the starred state to other states summing over phonon modes. (c) and (d) are color maps of coupling strength from the starred state to the lowest four exciton bands in the full BZ summed over exciton states at a given $\vQ$ and phonon modes. (e) and (f) show the normalized contribution to the scattering rate (denoted also by $\Gamma$ here) of the starred state from different phonon momentum \vq. Arrows in (a), (b), (c), and (d) highlight the region of large contribution to the scattering rate (which in general can be different from the magnitude of $G$) as shown in (e) and (f). (a) and (b) show strong state-selective couplings due to spin structure of the excitons.}
\label{fig:analysis}
\end{figure*}

We show in Fig. 1 (a) the calculated exciton dispersion (or exciton band structure) and a color map of the exciton relaxation times due to phonons in monolayer MoS$_2$ at 300K. In our calculation and consistent with previous results~\cite{Qiu2015}, the lowest-energy exciton has a COM momentum $\vQ=K$ while the lowest exciton with momentum $\vQ=\Gamma$ is 1.7 meV higher in energy (Both lowest-energy excitons are optically inactive). Consequently, the lowest energy $\vQ=K$ exciton has a longer relaxation time or lifetime, which is about 100 fs, since the available phase space for exciton-phonon scattering is limited at the lowest exciton band edge. However, to our surprise, the first optically bright exciton, the zero COM A exciton, also has a long lifetime comparable with the lowest $\vQ=K$ dark exciton, while the lowest energy dark exciton at $\Gamma$ has a much shorter lifetime. This is counterintuitive as one would expect the available scattering phase space for the A exciton to be larger than the lowest energy dark exciton with the same COM momentum.

To understand this result, we analyze the momentum and state-resolved exciton-phonon coupling matrix elements $G_{S'S\nu}(\vQ,\vq)$ in Fig.~\ref{fig:analysis}. In Fig.~\ref{fig:analysis} (a) and (b) we show color maps of the absolute value of $G_{S'S\nu}(\vQ,\vq)$ from the lowest-energy $\vQ=\Gamma$ exciton (blue-starred) and the A exciton (red-starred), respectively, to other exciton states summing over all phonon modes. We observe that both excitons are strongly coupled to $\vQ=M$ excitons in bands with the same spin configurations while the coupling strength is zero to excitons in other bands with a different internal spin structure, which is well-defined for states around $M$ and $K$. For example, the A exciton does not couple to the lowest two $\vQ=M$ excitons and the lowest-energy $\vQ=K$ exciton. This highly selective spin-configuration-dependent exciton-phonon coupling plays a crucial role in determining the relaxation time of the A exciton. In Fig.~\ref{fig:analysis} (c) and (d) we show color maps of exciton-phonon coupling from the starred states in Fig.~\ref{fig:analysis} (a) and (b) to exciton of different $\vQ$ summing over all phonon modes and the four lowest bands of final exciton states. These maps together with Fig. 2 (a) and (b) further demonstrate that the strongest coupling is the coupling of bands with the same configuration from $\vQ=\Gamma$ to $\vQ=M$. However, since excitons of the same spin configuration around $\vQ=M$ as the A exciton lay about 70 meV higher in energy, which is larger than the highest phonon energy (60 meV) in MoS$_2$, they are not able to contribute to the scattering rate. Because of the Bose factor, we find that the major contributions come from low-energy LA and TA modes at 300 K. In contrast, the excitons near the exciton band edge with $\vQ=M$ are energetically close to the lowest exciton state with $\vQ=\Gamma$ so the coupling between these two states dominate the contribution to the large scattering rate seen in Fig.~\ref{fig:Relax}. In Fig.~\ref{fig:analysis} (e) and (f) we show the phonon momentum $\vq$-resolved contribution $\Gamma_{\mathbf{q}}$ to the total scattering rates of the starred states in Fig.~\ref{fig:analysis} (a) and (b), respectively. We find that, indeed for the blue-starred exciton state, phonons with momentum close to $M$ contribute most to the scattering rate while phonons with momentum near $K$ and $\Gamma$ both have smaller contribution. On the other hand the A exciton mostly scatters to other exciton states via phonons with $\vq\sim K$ while $\vq\sim 0$ and $\vq\sim \Lambda_{min}$ also contribute. We note that the ring shape seen in Fig.~\ref{fig:analysis} (e) and (f) is due to the mismatch in the velocity of the exciton and phonon bands near the band edge, which limits the scattering density of final states.

\begin{figure*}[t]
     \centering
     \includegraphics[width=.8\textwidth]{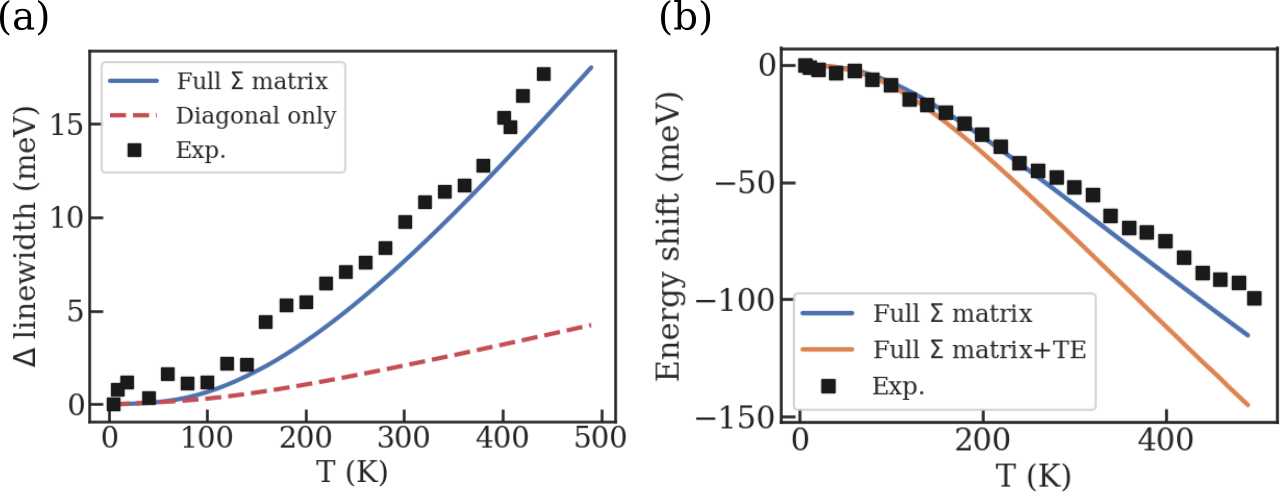}
\caption{(a) Temperature dependence of the linewidth computed with full exciton self-energy matrix (blue line) and diagonal matrix elements only (red dashed line) as compared to the experiment results from Ref.~\cite{Dey2016}. A temperature-independent constant corresponding to the measured low-temperature linewidth is subtracted from the experimental data. (b) Shift of the absorption peak as a function of temperature calculated with the full self-energy matrix (blue line) is compared with the results from same experiment in (a). Energy shift due to lattice thermal expansion (TE) effects are included and shown in the orange line.}
\label{fig:abs}
\end{figure*}

Experimentally, exciton-phonon interactions can be accessed by measuring the absorption linewidth and its temperature dependence. The contribution of exciton-phonon interactions to the optical absorption spectrum can be studied within the semiconductor Bloch equation~\cite{Koch2000,Rossi2002,Selig2016,Christiansen2017} or from the second order perturbation theory~\cite{Hopfield1961,Toyozawa1964,Segall1968,Perebeinos2005}. The derivation of the former and connection to the latter is given in the supplementary materials~\cite{SM}. We summarize the equation of motion of the exciton polarization which is directly related to the optical response,
\begin{align}
&\frac{dP^S_{\mathbf{0}}(t)}{dt} \nonumber \\
&=\frac{1}{i\hbar}\sum_{\lambda}\left(E_{S\mathbf{0}}\delta_{S\lambda}-\Sigma_{S\lambda\mathbf{0}}(\omega=E_{\lambda\mathbf{0}})\right)P^\lambda_{\mathbf{0}}\nonumber\\
&-\frac{1}{i\hbar}e\mathbf{E}(t)\cdot\mathbf{\Omega}^S,
\label{Eq:eom}
\end{align}
where $P^S_{\mathbf{0}}\equiv P^S_{\vQ=0}=\sum_{cv\vk} \psi^{S*}_{c\vk v\vk}\langle c^\dagger_{v\vk}c_{c\vk} \rangle$ is the exciton polarization with zero COM momentum, $\psi^S_{c\vk v\vk}$ is the exciton envelope function of a state $S$, and $c^\dagger_{v\vk}$ ($c_{c\vk}$) is the creation (annihilation) operator for an valence electron $v$ (a conduction electron $c$) with crystal momentum $\vk$. The second term in the right-hand side of Eq.~\ref{Eq:eom} describes the coupling to an external field $\mathbf{E}(t)$ within the dipole approximation and the exciton dipole operator is defined by $\mathbf{\Omega}^S\equiv \sum_{cv\mathbf{k}}d_{cv\mathbf{k}}\psi^{S*}_{cv\mathbf{k}}$ with the interband optical transition matrix element $d_{cv\mathbf{k}}$ given in the Kohn-Sham basis~\cite{SM}. Finally, the second quantity in the parenthesis for the sum in Eq.~\ref{Eq:eom} is the exciton self-energy expressed in the unperturbed exciton basis. It is a more complete expression for the self-energy and a matrix generalization of Eq.~\ref{Eq:sediag},
\begin{align}
\Sigma_{S\lambda,\mathbf{0}}(\omega)=\frac{1}{N_\mathbf{q}}&\sum_{n\nu\vq}G^*_{n S \nu}(0,\vq)G_{n \lambda \nu}(0,\vq) \nonumber \\ &\times\frac{N_{\nu\vq} + \frac{1}{2}\pm\frac{1}{2}}{\omega-E_{n\vq}\mp\omega_{\nu\vq}+i\gamma}.
\label{Eq:offse}
\end{align}
If the off-diagonal elements (i.e. the $\lambda$ not equal to S terms) of Eq.~\ref{Eq:offse} are ignored in Eq.~\ref{Eq:eom}, the computed absorption spectrum is a superposition of Lorentzian functions centered at exciton energies with linewidth equivalent to the relaxation time computed by taking the imaginary part of Eq.~\ref{Eq:sediag}. This approximation (we shall call it the diagonal approximation) is widely used in the literature~\cite{Antonius2017,Cannuccia2019,Chen2020,Reichardt2020}. The slope of the temperature dependence of the linewidth is often taken as a measure of the exciton-phonon coupling strength. In Fig.~\ref{fig:abs} (a) we plot the temperature dependence of the calculated linewidth using the diagonal approximation as the dashed red line. Within the diagonal approximation, the linewidth is severely underestimated as compared with the experimental results from Ref.~\cite{Dey2016}. The computed linewidth at 300 K is less than 2 meV which is about 3 times smaller than the experimental value. On the other hand, when the linewidth is evaluated at $\omega=E_\lambda$, using the full self-energy matrix expression given by Eq.~\ref{Eq:offse}, the results agree very well with experiment. Not only is the magnitude closer to experiment but the overall temperature dependence also agrees. We emphasize that the same phonon-induced self-energy matrix elements are used in both calculations and that only when using the full self-energy matrix, then we are able to obtain good agreement with experiment. 

\begin{figure*}[tb]
    \centering
    \includegraphics[scale=0.50]{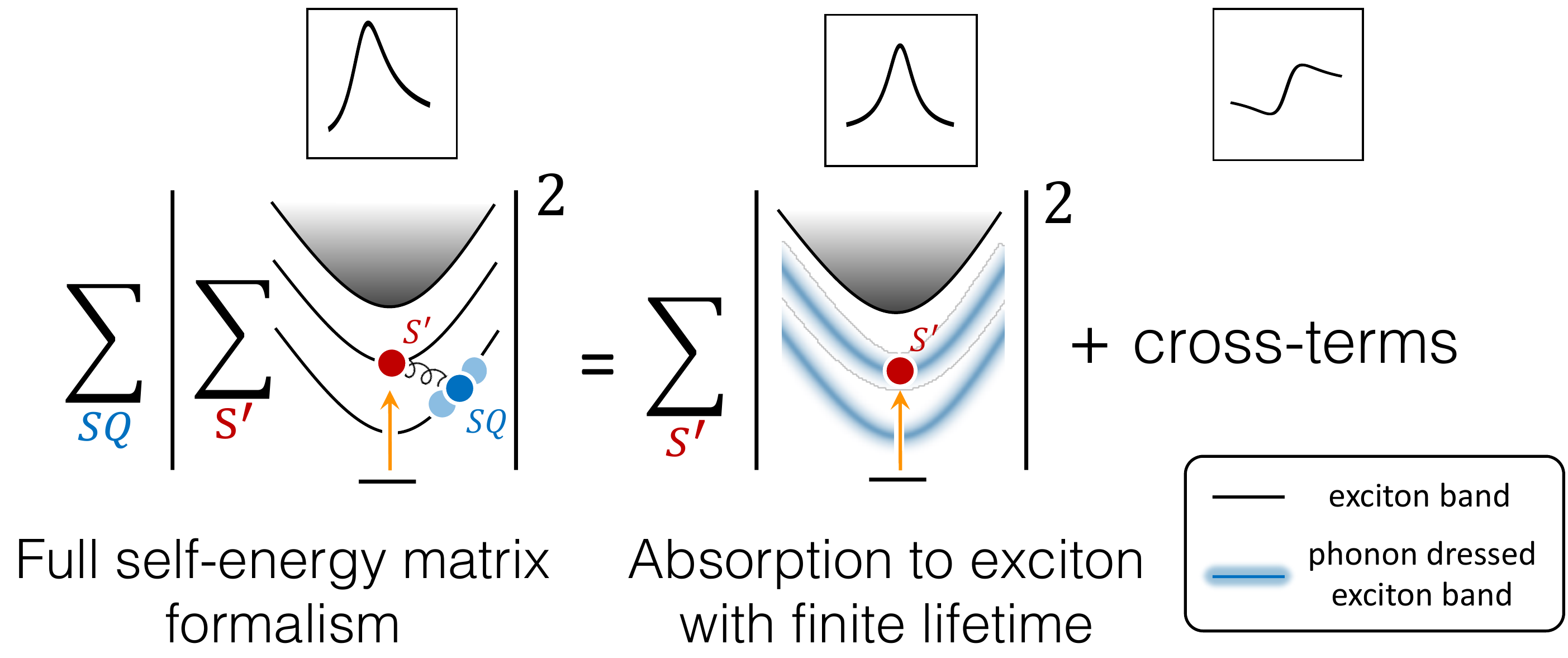}
    \caption{Computing $\varepsilon_2(\omega,T)$ with the full exciton self-energy matrix, $\Sigma_{SS'}(\omega)$, captures all the physics contained in a second order phonon-assisted absorption process, where a photon is first absorbed creating a virtual exciton, S’ (red dots), which subsequently scatters off a phonon to some final state S$\mathbf{Q}$ (blue dot). Expanding the innermost sum and subsequently taking the square modulus, two types of terms arise, those diagonal in $S'$ and those involve cross-terms. The former gives rise to a symmetric Lorentzian line shape and physically stems from photo-exciting to a finite lifetime quasiparticle band with finite energy broadening. The cross-terms are typically not included in previous \textit{ab initio} calculations but can be important and give rise to all the asymmetry seen in the total line shape.}
    \label{fig:model_abs}
\end{figure*}

The importance of including these off-diagonal self-energy terms and their effects on the exciton lineshape has been discussed at length by Toyozawa~\cite{Toyozawa1964}, Hopfield~\cite{Hopfield1961}, and others in model studies over 50 years ago; however, with rare exceptions~\cite{Schweiner2016}, such terms to our knowledge have not been included in contemporary calculations and never before from first principles. To map to Toyozawa's formalism, we take the external field to be monochromatic, $\mathbf{E}(t)=\mathbf{E}_0e^{i\omega t}$, and Fourier transform Eq. 2, solve for $\epsilon_2(\omega) = P^{tot}(\omega)/ \epsilon_0 E$ and explicitly rewrite the full matrix self-energy in terms of the exciton-phonon coupling matrix elements with the aid of Eq. \ref{Eq:offse} to find:
\begin{align}
&\varepsilon_2(\omega,T)  \nonumber \\
&\sim\frac{1}{N_\mathbf{q}}\sum_{S \mathbf{q} \nu} \bigg| \hat{\mathbf{e}} \cdot \sum_{S'} \frac{G_{SS'\nu}(0,\mathbf{q}) \mathbf{\Omega}_{S'}}{\omega - \tilde{E}_{S'}} \bigg|^2 \nonumber \\
&\times\bigg( \frac{1}{2}\pm\frac{1}{2}+N_{\nu\vq} \bigg)\delta(\omega-E_{S\mathbf{q}} \mp \omega_{\nu \mathbf{q}} ),
\label{Eq:indabs}
\end{align}
where $\hat{\mathbf{e}} = \mathbf{E}_0/|\mathbf{E}_0|$ is the polarization unit vector and $\tilde{E}_{S} = E_{S} + {\rm Re} \Sigma_S+i\Gamma_S$ with ${\rm Re} \Sigma_S$ and $\Gamma_S ={\rm Im}\Sigma_S$ denoting the real and imaginary part of the diagonal exciton self-energy defined in Eq.~\ref{Eq:sediag} respectively. More explicit details connecting Eqs~\ref{Eq:eom} and~\ref{Eq:indabs} are given in the SI.  An advantage of this expression over the semiconductor Bloch expression is that it is readily interpreted as describing a two step process within second order perturbation theory where a photon is absorbed and a phonon subsequently scatters off the excitation. We depict this process in Fig.~\ref{fig:model_abs}. Explicitly writing out the innermost sum on $S'$ in Eq.~\ref{Eq:indabs} and taking its square modulus, we can regroup the resulting terms based on whether they are diagonal or off-diagonal in the $S'$ index. Retaining only terms diagonal in $S'$ we arrive at what we call the diagonal approximation for $\varepsilon_2(\omega,T)$,
\begin{widetext}
\begin{align}
\varepsilon^{\rm diag}_2(\omega,T) &\sim \frac{1}{N_\mathbf{q}}\sum_{SS' \mathbf{q} \nu} \bigg| \hat{\mathbf{e}} \cdot \frac{G_{SS'\nu}(0,\mathbf{q}) \mathbf{\Omega}_{S'}}{\omega - \tilde{E}_{S'}} \bigg|^2 \bigg( \frac{1}{2}\pm\frac{1}{2}+N_{\nu\vq} \bigg)\delta(\omega-E_{S\mathbf{q}} \mp \omega_{\nu \mathbf{q}} ),
\label{Eq:diag_eps} \nonumber \\
&\approx \sum_{S'} |\hat{\mathbf{e}} \cdot \mathbf{\Omega}_{S'}|^2 \frac{\Gamma_{S'}}{(\omega-E_{S'}-{\rm Re}\Sigma_{S'})^2 + \Gamma_{S'}^2},
\end{align}
\end{widetext}
where we identify $\Gamma_S ={\rm Im}\Sigma_{SS\mathbf{0}}$ with the diagonal part of the self-energy defined in Eq.~\ref{Eq:sediag}. We see in this diagonal limit we recover the familiar Lorentzian lineshape with a full width at half maximum given by the imaginary part of the diagonal element of the exciton self-energy. This symmetric lineshape is a familiar spectroscopic signature stemming from the process of photo-exciting to a transient excitonic state with some finite lifetime. The cross-terms neglected in passing from Eq.~\ref{Eq:indabs} to Eq.~\ref{Eq:diag_eps} are in some cases very important and give rise to asymmetric contributions to the exciton lineshape (see the Supplemental Information for detail). To summarize the above discussion, in Fig.~\ref{fig:model_abs} we give a pictorial representation of the relation between the full and diagonal approximations of the exciton self-energy for $\varepsilon_2(\omega,T)$.

In Fig.~\ref{fig:abs} (b) we further show the temperature dependence of the shift of the A exciton peak, which represents the renormalization of exciton energy due to exciton-phonon couplings. Again we find remarkably good agreement between the calculation and the extracted experimental values. The computed energy shift is around 60 meV at 300 K, which is comparable to the band gap renormalization due to electron-phonon couplings~\cite{Molina2016}. However, when the effects from lattice expansion~\cite{Huang2014} are included the results deviate from the experiments as shown by the blue line. We expect that the Debye-Waller term which is not considered in this work could partially compensate the lattice thermal expansion~\cite{Gab2015} effect.

In conclusion, our first-principle theory and calculation of the exciton-phonon coupling strength in a monolayer MoS$_2$ shows highly selective nature of exciton-phonon couplings between different exciton states due to the internal spin structure of the excitons. In particular, we have shown that the A exciton has surprisingly long life time about 100 fs at 300 K, which is longer than the first $Q=0$ dark exciton state at the exciton band edge. Although it has strong couplings with excitons around $M$ in the same band, the large difference in energy limits the available scattering phase space. Hence the dominant contribution comes from the relatively weak couplings to the $K$ excitons. Moreover, we show that the temperature-dependent absorption linewidth and peak energy shift computed with the full exciton self-energy matrix agrees much better with the experimental data, which emphasize the importance of the interference effects among inter-exciton band transitions. The theory also leads to an asymmetric lineshape when inter-exciton couplings are strong, which maybe an interesting feature worthy of further investigation in experiment.  

We thank Prof. Denis Karaiskaj for kindly providing the absorption spectrum of MoS$_2$ for our analysis. This work was primarily supported by the Center for Computational Study of Excited State Phenomena in Energy Materials (C2SEPEM), funded by the U.S. Department of Energy, Office of Science, Basic Energy Sciences, Materials Sciences and Engineering Division under Contract No. DE-AC02-05CH11231, as part of the Computational Materials Sciences Program which provided the method, algorithm and code developments as well as GW-BSE calculations. The many-body second-order perturbation theory and analysis for the optical response was supported by the Theory of Materials Program at LBNL which is funded by the U.S Department of Energy, Office of Science, Basic Energy Sciences, Materials Sciences and Engineering Division under Contract No. DE-AC02-05CH11231. Y.-H. C. was supported by the National Science and Technology Council of Taiwan under grant no. 110-2124-M-002-012. We acknowledge the use of computational resources at the National Energy Research Scientific Computing Center (NERSC), a DOE Office of Science User Facility supported by the Office of Science of the U.S. Department of Energy under Contract No. DE-AC02-05CH11231, and National Center for High-performance Computing (NCHC) in Taiwan. The authors acknowledge the Texas Advanced Computing Center (TACC) at The University of Texas at Austin for providing HPC resources that have contributed to the research results reported within this paper.

\bibliography{ref}

\end{document}